\documentclass[12pt]{article}
\usepackage{amssymb}
\newcommand{\beq}{\begin{equation} }
\newcommand{\eeq} {\end{equation} }

\newcommand{\bed}{\begin{displaymath}}
\newcommand{\eed}{\end{displaymath}}
\begin{document}

\begin{titlepage}
\begin{flushright}
OSU-HEP-03-8\\
May 2003\\
\end{flushright}
\vskip 2cm
\begin{center}
{\large\bf \textbf Gauged Baryon Parity and Nucleon Stability}
\vskip 1cm {\normalsize\bf
K.S.\ Babu\footnote{E-mail address: babu@okstate.edu}, Ilia
Gogoladze\footnote{ On leave of absence from: Andronikashvili
Institute of Physics, GAS, 380077 Tbilisi, Georgia.  \\ E-mail
address: ilia@hep.phy.okstate.edu}  and
Kai Wang\footnote{E-mail address: wangkai@hep.phy.okstate.edu}} \\
\vskip 0.5cm
{\it Department of Physics, Oklahoma State University\\
Stillwater, OK~~74078-3072, USA\\ [0.1truecm] }

\end{center}

\begin{abstract}
We show that the Standard Model Lagrangian, including small
neutrino masses, has an anomaly-free discrete $Z_6$ symmetry. This
symmetry can emerge naturally from $(I^{3}_{R}+L_i+L_j-2L_k)$
gauge symmetry ($L_i$ is the $i$th lepton number) and ensure the
stability of the nucleon even when the threshold of new physics
$\Lambda$ is low. All $\Delta B=1$ and $\Delta B=2$ ($B$ is the
baryon number) effective operators are forbidden by the $Z_6$
symmetry. $\Delta B=3$ operators are allowed, but they arise only
at dimension 15. We estimate the lifetime for ``triple nucleon
decay" resulting from these operators and find that $\Lambda$ can
be as low as $10^2~ \mathrm{GeV}$. We suggest a simple mechanism
for realizing reasonable neutrino masses and mixings even with
such a low scale for $\Lambda$.

\end{abstract}

\end{titlepage}

\section{Introduction}

The Standard Model (SM) has been highly successful in explaining
all experimental observations in the energy regime up to a few
hundred GeV. However, it is believed to be an effective field
theory valid only up to a cutoff scale $\Lambda$.
Non-renormalizable operators which are gauge invariant but
suppressed by appropriate inverse powers of $\Lambda$ should then
be considered in the low energy effective theory \cite{operators}.
The dimension 5 operator $\ell\ell H H/{\Lambda_L}$ ($\ell$ is the
lepton doublet) which violates lepton number ($L$) by two units is
the lowest dimensional of such operators. Experimental evidence
for neutrino masses suggests the effective scale of $L$-violation
is around $\Lambda_L\sim 10^{14}-10^{15}~\mathrm{GeV}$. The $d=6$
operator $QQQ\ell/{ {\Lambda}^2_B}$ violates both baryon number
($B$) and lepton number and leads to the decay of the nucleon. The
current limits on proton lifetime are $\tau_p
>5\times 10^{33}~\mathrm{yrs}$ for $p\rightarrow e^+ \pi^{0}$ and
$\tau_p
>1.6\times 10^{33}~\mathrm{yrs}$ for $p\rightarrow \bar{\nu}K^{+}$
\cite{pdg}. These limits imply that
$\Lambda_B>10^{15}~\mathrm{GeV}$. Grand Unified Theories with or
without supersymmetry generate such $B$-violating operator with
$\Lambda_B\sim 10^{14}-10^{16}~\mathrm{GeV}.$ These theories are
currently being tested through nucleon decay. Any new physics with
a threshold $\Lambda$ less than the GUT scale will thus be
constrained by both proton lifetime and neutrino masses
\cite{operators, zee, leung}.

It is widely anticipated that new physics will show up around the
TeV scale. This expectation is based on the stability of the Higgs
mass. Supersymmetric theories with threshold of a few hundred GeV
have been extensively studied as a solution to the Higgs mass
problem (or the hierarchy problem). Indeed, these theories allow
for both $B$ and $L$ violating operators. Care must be (and
usually is) taken to satisfy the experimental constraints.

A second scenario that has been widely discussed recently to
address the hierarchy problem is the possible existence of large
extra dimensions \cite{add}. Here the fundamental scale of gravity
$M_{\rm Pl}$ is lowered from its 4 dimensional value of
$10^{19}~\mathrm{GeV}$ to $M_*\sim$ a few TeV. Non-renormalizable
operators associated with quantum gravity which are only
suppressed by inverse power of $M_*$ will then be present in the
low energy theory. $L$ and $B$ violation would put major
constraints on these theories. Similar remarks will apply to other
solutions to the hierarchy problem, such as composite Higgs,
little Higgs \cite{littlehiggs}, etc.

In this paper we address the role discrete gauge symmetries
\cite{kw} can play in suppressing baryon number and lepton number
violating processes even when the scale of new physics is low. The
SM effective lagrangian does not have a continuous anomaly-free
symmetry that can suppress these processes. This is our reason for
focusing on discrete symmetries. It is preferable that such
symmetries have a gauge origin since all global symmetries are
expected to be violated by the quantum gravitational effects
\cite{hawking}. We find that the SM lagrangian, including small
neutrino masses, has an anomaly-free $Z_6$ symmetry. This $Z_6$
acts as discrete baryon number and forbids potentially dangerous
$\Delta B=1$ and $\Delta B=2$ nucleon decay processes. $\Delta
B=3$ operators are allowed, but they arise only as $d=15$
operators. ``Triple nucleon decay" can proceed through these
operators; we estimate its rate and find $\Lambda_B$ can be as low
as $10^2~\mathrm{GeV}$. We show that the $Z_6$ symmetry has a
natural gauge origin in the anomaly-free $U(1)$ corresponding to
$I^{3}_{R}+L_i+L_j-2L_k$, where $L_i$ is $i$th lepton number and
$I^3_R$ is the third component of the righthanded isospin. No new
particles are introduced to cancel anomalies except the
righthanded neutrinos. We show how an additional $Z_5$ or $Z_7$
gauge symmetry generates the desired neutrino masses and mixings
even with a low scale for $\Lambda$.

Discrete gauge symmetries have been utilized in suppressing
nucleon decay \cite{ibanez} as well as in addressing other aspects
of physics such as solving the $\mu$ problem of supersymmetry,
fermion mass hierarchy problem \cite{wk, axion} and the stability
of the axion \cite{axion}. A $Z_3$ baryon parity was found in Ref.
\cite{ibanez} that suppresses nucleon decay. In order for it to
have a gauge origin, extra particles were introduced.\footnote{The
charges of the SM particles under the $U(1)$ symmetry in Ref.
\cite{ibanez} are $\{Q,~u~,~d,~l,~e\}=(0,-1,1,-1,-1)$. The model
has extra leptons $l'(3)$,
 $\bar{l}'(0)$, $\bar{e}'(3)$, $\bar{e}'(0)$ (where the $U(1)$ charges are given in parentheses) as well as  exotic leptons with electric charge of
 $\pm 1/3$ and $0$: $2\times F^{1/3}(3)$, $1\times F^{-1/3}(3)$, $2\times F^{-1/3}(0)$, $1\times F^{1/3}(0)$  and $5\times N^{0}(-3)$, $N^{0}(1)$, $N^{0}(3)$.} There have been also attempts to gauge baryon number
by introducing a fourth family of fermions \cite{carone}, or with
the use of extra dimensions \cite{mohapatra}. The spirit of this
paper is similar to that of Ref. \cite{ibanez, carone, mohapatra},
our models are however much more economical.

\section{Gauged Baryon Parity and Suppression of Proton Decay}

In this section we show that the SM lagrangian has a discrete
$Z_6$ gauge symmetry which forbids all $\Delta B=1$ and $\Delta
B=2$ baryon violating effective operators. This can be seen as
follows. The SM Yukawa couplings incorporating the seesaw
mechanism to generate small neutrino masses is \beq {\cal L}_{\rm
Y} = Qu^{c}H+Qd^{c}{H}^*+\ell e^{c}{H}^*+\ell\nu^{c}H+M_R
\nu^{c}\nu^{c}~~.\eeq Here we have used the standard (lefthanded)
notation for the fermion fields and have not displayed the Yukawa
couplings or the generation indices. This lagrangian respects a
$Z_6$ discrete symmetry with the charge assignment as shown in
Table \ref{z6charge}. Also shown in Table \ref{z6charge} are the
charge assignments under the $Z_3$ and $Z_2$ subgroups of $Z_6$.
The $Z_3$ assignment is identical to that in Ref. \cite{ibanez}
\begin{table}[ht]
 \begin{center}
  {\renewcommand{\arraystretch}{1.1}
 \begin{tabular}{|c| c c c c c c c | }
   \hline
  \rule[5mm]{0mm}{0pt} & $Q$ & $u^c$ & $d^c$ & $\ell$ & $e^c$ & $\nu^{c}$& $H$ \\
  \hline
  \rule[5mm]{0mm}{0pt}
 $Z_6$&6 & 5 & 1 & 2 & 5 &3 & 1  \\
 $Z_3$&3 & 2 & 1 & 2 & 2 &3 & 1  \\
 $Z_2$&2 & 1 & 1 & 2 & 1 &1 & 1  \\
     \hline
\end{tabular}
 }
  \caption{\footnotesize Family-independent $Z_6$ charge assignment of the SM fermions and the Higgs boson along with the charges under $Z_3$ and $Z_2$ subgroups.  }
  \label{z6charge}
 \end{center}
\end{table}

From Table \ref{z6charge} it is easy to calculate the $Z_6$
crossed anomaly coefficients with the SM gauge groups. We find the
$SU(3)_C$ or $SU(2)_L$ anomalies to be\begin{eqnarray}
A_{{[SU(3)_C]}^2\times Z_6}=3N_g\nonumber\\
A_{{[SU(2)_L]}^2\times Z_6}=N_g~
\end{eqnarray}where $N_g$ is the number of generations.
The condition for a $Z_N$ discrete group to be anomaly-free is
\beq A_i=\frac{N}{2}~~mod~N~\eeq where $i$ stands for $SU(3)_C$
and $SU(2)_L$. For $Z_6$, this condition reduces to $
A_i=3~mod~6$, so when $N_g=3$, $Z_6$ is anomaly-free. Obviously,
the $Z_3$ and $Z_2$ subgroups are also anomaly-free. The
significance of this result is that unknown quantum gravitational
effects will respect this $Z_6$. It is this feature that we
utilize to stabilize the nucleon. Absence of anomalies also
suggests that the $Z_6$ may have a simple gauge origin.

To see how the $Z_6$ forbids $\Delta B=1$ and $\Delta B=2$
processes, we note that it is a subgroup of $U(1)_{2Y-B+3L}$ where
$Y$ is SM hypercharge \cite{wk}. We list in Table \ref{charge} the
charges under the three $U(1)$ symmetries. Comparing the last line
of Table \ref{charge} with the charge assignment of Table
\ref{z6charge} it is clear that the $Z_6$ is a subgroup of
$U(1)_{2Y-B+3L}$. Any $Z_6$ invariant effective operator must then
satisfy \beq \label{zz} 2\Delta Y-\Delta B+3\Delta L=0~~mod~6.\eeq
Invariance under $U(1)_Y$ implies $\Delta Y=0$. Consider $\Delta
B=1$ effective operators which must then obey (from Eq.
(\ref{zz})) $3\Delta L=1~~mod~6$. This has no solution, since
$3\Delta L=0~~mod~3$ from Table \ref{charge}. Similarly, $\Delta
B=2$ operators must obey $3\Delta L=2~~mod~6$ which also has no
solution. $\Delta B=3$ operators, which corresponds to $3\Delta
L=0~~mod~6$, are allowed by this $Z_6$. Such operators have
dimension 15 or higher and have suppression factors of at least
$\Lambda^{-11}$. These will lead to ``triple nucleon decay"
processes where three nucleons in a heavy nucleus undergo
collective decays leading to processes such as $pnn\rightarrow
e^+\pi^0$. We estimate the rates for such decay in Section 4 and
find that $\Lambda$ can be as low as $10^2~\mathrm{GeV}$.
\begin{table}[ht]
 \begin{center}
  {\renewcommand{\arraystretch}{1.1}
 \begin{tabular}{|c| c c c c c c c | }
   \hline
  \rule[5mm]{0mm}{0pt} & $Q$ & $u^c$ & $d^c$ & $\ell$ & $e^c$ & $\nu^{c}$& $H$ \\
  \hline
  \rule[5mm]{0mm}{0pt}
 $U(1)_B$& $1/3$&$-1/3$&$-1/3$&0&0&0&0\\
    \rule[5mm]{0mm}{0pt}
 $U(1)_L$& 0&0&0&$1$&$-1$&$-1$&0\\
    \rule[5mm]{0mm}{0pt}
 $U(1)_Y$&$1/6$&$-2/3$&$1/3$&$-1/2$&1&0&$1/2$\\
  \rule[5mm]{0mm}{0pt}
  $U(1)_{2Y-B+3L}$&0&$-1$&1&2&$-1$&$-3$&1\\
      \hline
\end{tabular}
 }
  \caption{\footnotesize Charge assignment under $B$, $L$, $Y$ and $U(1)_{2Y-B+3L}$ which contains the $Z_6$ of Table \ref{z6charge}.}
  \label{charge}
 \end{center}
\end{table}

\section{Embedding $Z_6$ in $I^{3}_{R}+L_i+L_j-2L_k$}

It is interesting to see if the $Z_6$ symmetry of Table
\ref{z6charge} can be realized as an unbroken subgroup of a gauged
$U(1)$ symmetry. Although the $Z_6$ is a subgroup of the
$U(1)_{(2Y-B+3L)}$, this $U(1)$ would be anomalous without
enlarging the particle content. We have found a simple and
economic embedding of $Z_6$ into a $U(1)$ gauge symmetry
associated with $I^{3}_{R}+L_i+L_j-2L_k$. Here $L_i$ is the $i$th
family lepton number and $i\neq j\neq k$. No new particles are
needed to cancel gauge anomalies. With the inclusion of
righthanded neutrinos $I^{3}_{R}=Y-(B-L)/2$ is an anomaly-free
symmetry. $L_i+L_j-2L_k$, which corresponds to the $\lambda_8$
generator acting in the leptonic $SU(3)$ family space, is also
anomaly-free.

The charges of the SM particles under this $U(1)$ are
\begin{displaymath}
\begin{array}{cccc}
 Q_{i}=(0,0,0),& {u_{i}}^c=(-1, -1, -1), & {d_{i}}^c=(1, 1,
1),& \\
 \ell_{i}=(-4,2, 2),& {e_{i}}^c=(5, -1, -1),& {\nu_{i}}^c=(3, -3,
-3)~, &H=1.\\
\end{array}%
\nonumber
\end{displaymath}This
charge assignment allows all quark masses and mixings as well as
charged lepton masses. When the $U(1)$ symmetry breaks
spontaneously down to $Z_6$ by the vacuum expectation value of a
SM singlet scalar field $\phi$ with a charge of 6, realistic
neutrino masses and mixings are also induced. The relevant
lagrangian for the righthanded neutrino Majorana masses is
\beq\label{rh}
 {\cal L}_{\rm
 \nu^c}=M_{12}\nu^c_1\nu^c_2+M_{13}\nu^c_1\nu^c_3+\nu^c_3\nu^c_3\phi+\nu^c_2\nu^c_2\phi+\nu^c_1\nu^c_1{\phi}^*
 +\nu^c_3\nu^c_2\phi~.
 \eeq
After integrating out the heavy righthanded neutrinos we obtain
the following $\Delta L=2$ effective operators: \beq
\label{bb}{\cal L}_{\Delta L=2}=\frac{1}{\Lambda}(\ell_1 \ell_2 H
H+\ell_1 \ell_3 H H+\ell_1 \ell_1 H H \epsilon+\ell_2 \ell_2 H
H{\epsilon}^{*}+\ell_2 \ell_3 H H{\epsilon}^{*}+\ell_3 \ell_3 H
H{\epsilon}^{*}).\eeq Here $\Lambda\sim M_{12}\sim M_{13}$ is the
scale of $L$-violation and we have defined
$\epsilon\equiv\left<\phi\right>/\Lambda$. For $\epsilon\ll 1$,
this lagrangian leads to the inverted mass hierarchy pattern for
the neutrinos which is well consistent with the current neutrino
oscillation data. This neutrino mass mixing pattern is analogous
to the one obtained from $L_e-L_\mu-L_\tau$ symmetry \cite{babu}.
However, here the $U(1)$ is a true gauge symmetry.

We have also investigated other possible $U(1)$ origin of the
$Z_6$ symmetry and found the $I^3_R+L_i+L_j-2L_k$ combination to
be essentially unique. To see this, let us assign a general
$U(1)_X$ charge for $i$th generation of the SM fermions consistent
with the $Z_6$ symmetry as \bed
\{Q_i,u_i^c,d_i^c,\ell_i,e_i^c,\nu_i^{c}\}
=\{6m^{(i)}_1,5+6m^{(i)}_4,1+6m^{(i)}_3,2+6m^{(i)}_2,5+6m^{(i)}_5,3+6m^{(i)}_6\}\eed
where $m^{(i)}_j$ are all integers. The Higgs field has a charge
$H=1+6m_0$. If we impose the invariance of the Yukawa couplings of
the charged fermions and Dirac neutrinos for each generation, the
anomaly coefficients from $i$th generation become
\begin{eqnarray}\label{anomaly}
A^{(i)}_{{[SU(3)_C]}^2\times U(1)_X}&=&0\nonumber\\
A^{(i)}_{{[SU(2)_L]}^2\times U(1)_X}&=& 1+9m^{(i)}_1+3m^{(i)}_2\nonumber\\
A^{(i)}_{{[U(1)_Y]}^2\times U(1)_X}&=& -(1+9m^{(i)}_1+3m^{(i)}_2)\nonumber\\
A^{(i)}_{{[U(1)_X]}^2\times U(1)_Y}&=& [5+m_0] A^{(i)}_{{[SU(2)_L]}^2\times U(1)_X}\nonumber\\
A^{(i)}_{{[U(1)_X]}^3}&=& [5+m_0]^{2}A^{(i)}_{{[SU(2)_L]}^2\times
U(1)_X}.
\end{eqnarray}
The coefficient for the mixed gravitational anomaly for each
generation is zero. From Eq. (\ref{anomaly}), it follows that
$A_2=\sum_i A^{(i)}_{{[SU(2)_L]}^2\times U(1)_X}=
\sum_i(1+9m^{(i)}_1+3m^{(i)}_2)=0$ can be satisfied only when all
three generation contributions are included. Once $A_2=0$ is
satisfied, all other anomaly coefficients will automatically
vanish. $A_2=0$ can be rewritten in a familiar form as $3\sum_i
Q_i+\sum_i \ell_i=0$. Thus we see that any $U(1)$ symmetry
satisfying this condition and consistent with the $Z_6$ charge
assignment can be a possible source of $Z_6$. If the $Q_i$ are
different for different generations, quark mixings cannot be
generated without additional particles. By making a shift
proportional to hypercharge, we can set $Q_i=0$ for all $i$. Two
obvious solutions to $\sum_i \ell_i=0$ are $\ell_i=(1,1,-2)$ and
$\ell_i=(1,-1,0)$. The latter one does not reproduce the $Z_6$
charge assignment while the former one does, which is our solution
when $I^3_R$ is added to it.

A related $B-3L_{\tau}$ has been discussed in Ref. \cite{ma}. This
is the same as $B-L$ plus $L_{e}+L_{\mu}-2L_{\tau}$. In Ref
\cite{ma}, only one righthanded neutrino $\nu^c_{\tau}$ is
introduced so the seesaw mechanism applies only for one light
neutrino. The other two neutrinos receive small masses from
radiative correction. In our model, since there are three
righthanded neutrinos, all the neutrino masses arise from the
conventional seesaw mechanism.

\section{$\Delta B=3$ Operators and Triple Nucleon Decay}

We now list the lowest dimensional (d=15) $\Delta B=3$ effective
operators which are consistent with the $Z_6$ symmetry. Imposing
gauge invariance and Lorentz invariance, we find them to be:
\begin{eqnarray}\label{op} &&~~\bar{u^c}^{4}\bar{d^c}^{5}\bar{e^c},
~~{\bar{u^c}}^{2}{\bar{d^c}}^{7}{e^c},
~~Q{\bar{u^c}}^{3}{\bar{d^c}}^{5}\ell,
~~Q{\bar{u^c}}^{2}{\bar{d^c}}^{6}{\bar{\ell}},
~~Q^{2}{\bar{u^c}}^{3}{\bar{d^c}}^{4}{\bar{e^c}},\nonumber\\&&
~~Q^{2}{\bar{u^c}}{\bar{d^c}}^{6}{e^c},
~~Q^{3}{\bar{u^c}}^{2}{\bar{d^c}}^{4}\ell,
~~Q^{3}{\bar{u^c}}{\bar{d^c}}^{5}{\bar{\ell}},
~~Q^{4}{\bar{u^c}}^{2}{\bar{d^c}}^{3}{\bar{e^c}},
~~Q^{4}{\bar{u^c}}{\bar{d^c}}^{4}{\nu^c},\nonumber\\&&
~~Q^{4}{\bar{d^c}}^{5}{e^c},~~Q^{5}{\bar{u^c}}{\bar{d^c}}^{3}\ell,
~~Q^{5}{\bar{d^c}}^{4}{\bar{\ell}},
~~Q^{6}{\bar{u^c}}{\bar{d^c}}^{2}{\bar{e^c}}, ~~
Q^{7}{\bar{d^c}}^{2}\ell, ~~Q^{8}{\bar{d^c}}{\bar{e^c}}~.
\end{eqnarray} Here Lorentz, gauge and flavor indices are suppressed.
These operators can lead to ``triple nucleon decay". The dominant
processes are
\begin{eqnarray}\label{decay}
ppp&\rightarrow& e^{+}+\pi^{+}+\pi^{+}\nonumber\\
ppn&\rightarrow& e^{+}+\pi^{+}\nonumber\\
pnn&\rightarrow& e^{+}+\pi^{0}\nonumber\\
nnn&\rightarrow& \bar{\nu}+\pi^{0}~.
\end{eqnarray}

Tritium ($^{3}H$) and Helium-3 ($^{3}He$) are examples of
three-nucleon systems in nature. These nuclei are unstable and
undergo $\beta$-decay with relatively short lifetime. In the
presence of operators of Eq. (\ref{op}), $^{3}H\rightarrow
e^{+}+\pi^{0}$ and $^3 He\rightarrow e^{+}+\pi^{+}$ decays can
occur. However, there is no stringent experimental limits arising
from these nuclei. So we focus on triple-nucleon decay in Oxygen
nucleus where there are experimental constraints from water
detectors. To estimate the decay lifetime we need to first convert
the nine-quark operators of Eq. (\ref{op}) into three-nucleon
operators and subsequently into Oxygen nucleus.

We choose a specific operator
$Q^5\bar{d^c}^4\bar{\ell}/\Lambda^{11}$ as an example to study the
process $pnn\rightarrow e^{+}+\pi^{0}$ triple nucleon decay
process. This induces the effective three-nucleon operator in
Oxygen nucleus \beq\label{aa}
\frac{Q^5\bar{d^c}^4\bar{\ell}}{{\Lambda}^{11}}\sim\frac{\beta^3
(1+D+F)}{\sqrt{2}f_{\pi}{\Lambda^{11}}}(\pi nnpe)~,\eeq where
$\beta\simeq 0.014~{GeV}^{3}$ is the matrix element to convert
three quarks into a nucleon \cite{jlqcd}. $F\simeq 0.47$, $D\simeq
0.80$ are chiral lagrangian factors, and  $f_{\pi}=139$ MeV is the
pion decay constant.

We now estimate the wave-function overlap factor for three
nucleons inside Oxygen nucleus to find each other. This is based
on a crude free fermi gas model where the nucleons are treated as
free particles inside an infinite potential well. A single nucleon
wave function is given by $\psi_{m}(x)=\sqrt{2/{r}}\sin({m\pi
x/{r}})$, where $r$ is the size of nucleus and $m$ is the energy
level. Incorporating isospin and Pauli exclusion principle, the
highest energy level which corresponds to $m=4$ is found to have 2
protons and 2 neutrons. We assume the highest level has the most
probability to form a Tritium-like ``bound state" of three
nucleons. The probability for three nucleons in Oxygen nucleus to
overlap in a range the size of Tritium nucleus is \beq
P\sim\frac{4}{3}\int^{\sqrt[3]{\frac{3}{16}}}_{0}d\left({\frac{x_1}{r}}\right)d\left({\frac{x_2}{r}}\right)d\left({\frac{x_3}{r}}\right){\left(\sin\left({\frac{4\pi
x_1}{r}}\right)\sin\left({\frac{4\pi
x_2}{r}}\right)\sin\left({\frac{4\pi x_3}{r}}\right)\right)}^2\sim
0.0253~,\eeq where $\sqrt[3]{\frac{3}{16}}$ is the ratio between
the radii of Tritium nucleus and Oxygen nucleus, since $R \propto
A^{1/3}$ (A is the atomic number). So the effective baryon number
violating operator of Eq. (\ref{aa}) becomes \beq
\frac{P\beta^{3}}{\sqrt{2}f_{\pi}{\Lambda^{11}}R^{3}}( {^{3}H}\pi
e )~.\eeq The triple nucleon decay lifetime can then be estimated
to be \beq \tau\sim \frac{16\pi{f_{\pi}}^2{\Lambda}^{22}R^{6}}{
P^{2}\beta^{6} M_{^{3}H}~}.\eeq By putting the current limit on
proton lifetime of $3\times 10^{33}~\mathrm{yrs}$, we obtain:\beq
\Lambda\sim 10^2~\mathrm{GeV}~.\eeq Thus we see the $Z_6$ symmetry
ensures the stability of the nucleon.

To test our crude model of nuclear transition, we have also
evaluated the double nucleon decay rate within the same approach
and found our results to be consistent with other more detailed
evaluations \cite{double}.

\section{Realistic Neutrino Masses with Low Threshold}

If the threshold of new physics is low, but is still much larger
than a TeV, small neutrino masses can be generated by the seesaw
mechanism as discussed in Section 3 through the effective
operators $\ell\ell H H/\Lambda$. If the threshold is as low as a
few TeV, the induced neutrino mass will be too large. Here we show
a mechanism by which such operators can be suppressed by making
use of a discrete $Z_N$ symmetry (with $N$ odd) surviving to low
scale. This $Z_N$ has a natural embedding in $B-L$ gauge symmetry.

Consider the following effective operators in the low energy
lagrangian: \beq\label{cc} {\cal L}\supset\ell\ell H
H\frac{S^{6}}{\Lambda^{7}}+\frac{S^{2N}}{{ \Lambda^{2N-4}}}~.\eeq
Here $S$ is a singlet field which has charge $(1,~3)$ under
$Z_N\times Z_6$ while $\ell$ has charge $(-3,~2)$. (The $Z_6$
charges of SM particles are as listed in Table \ref{z6charge}.)
The first term in Eq. (\ref{cc}) respects a $U(1)$ symmetry while
the second term reduces this to $Z_6\times Z_N$. If $S$ develops a
VEV of order $10^2~\mathrm{GeV}$, realistic neutrino mass can
arise even when $\Lambda$ is low. For example, if
$\Lambda=10~\mathrm{TeV}$ and $S=10^2~\mathrm{GeV}$, the neutrino
mass is of order $v^2 \left<S\right>^6/{\Lambda^{7}}\sim
0.4~\mathrm{eV}$, which is consistent with the mass scale
suggested by the atmospheric neutrino oscillation data.

Two explicit examples of the $Z_N$ symmetry with $N=5$ and $7$ are
shown in Table \ref{nu}. These $Z_N$ symmetries are free from
gauge anomalies. In the $Z_5$ example, the crossed anomaly
coefficients for $SU(3)_C$ and $SU(2)_L$  are $5N_g$ and $5N_g/2$
respectively showing that $Z_5$ is indeed anomaly-free. For $Z_7$,
these coefficients are $7N_g$ and $7N_g/2$, so it is also
anomaly-free.
\begin{table}[ht]\label{nu}
 \begin{center}
  {\renewcommand{\arraystretch}{1.1}
  \begin{tabular}{|c| c  c  c  c c   cc |   }
     \hline
     \rule[5mm]{0mm}{0pt}
   Field & $Q$ & $u^{c}$ &
   $d^{c}$ & $\ell$ &
   $e^{c}$  &
   $H$  & $S$\\
   \hline
   $Z_5$ & 1 & 4 & 4 &
   2 & 3  & 0  & 1\\
   $Z_7$ & $1$ & 6 & 6 &
   4 & 3  & 0  & 1\\
   \hline
  \end{tabular}
  }
  \caption{\footnotesize $Z_N$ charge assignment for $N=5$ and $7$.}
  \label{z7}
 \end{center}
\end{table}

It is interesting to ask if the $Z_N$ can be embedded into a
gauged $U(1)$ symmetry. A simple possibility we have found is  to
embed this $Z_N$ into the anomalous $U(1)_A$ symmetry of string
origin with the anomalies cancelled by the Green-Schwarz mechanism
\cite{gs}. Consider $U(1)_{B-L}$ without the right handed
neutrinos but with the inclusion of vector-like fermions which
have the quantum numbers of ${\bf 5}(3)$ and ${\bar{\bf 5}}(2)$
under $SU(5)\times U(1)_A$. This $U(1)_A$ is anomaly-free by
virtue of the Green-Schwarz mechanism. When this $U(1)_A$ breaks
down to $Z_5$, the extra particles get heavy mass and are removed
from the low energy theory which is the $Z_6\times Z_5$ model.

Without the second term in Eq. (\ref{cc}), the phase of $S$ field
will be massless upon spontaneous symmetry breaking. This Majoron
field \cite{majoron} would however acquire a mass from the second
term of Eq. (\ref{cc}). In the $Z_6\times Z_5$ model, the mass of
the Majoron is of order $\left<S\right>^7/\Lambda^6\sim
100~\mathrm{keV}$. In the $Z_6\times Z_7$ model, the Majoron mass
is of order $\left<S\right>^{11}/\Lambda^{10}\sim 10~\mathrm{eV}$.
Such a Majoron with a mass of either $100~\mathrm{keV}$ or
$10~\mathrm{eV}$ is fully consistent with constraints from early
universe cosmology \cite{babumj}. The interaction term $\ell\ell H
H S^6/{\rm \Lambda}^7$ induces the Majoron decay $S\rightarrow \nu
\nu$ with a Yukawa coupling
$Y_{S\rightarrow\nu\nu}=6m_\nu/\left<S\right>\sim 10^{-11}$. The
decay rate of the Majoron can be estimated to be\beq
\Gamma=\frac{Y^2_{S\rightarrow\nu\nu}m_S}{8\pi}\sim 10^{-23}
m_S~.\eeq This corresponds to a Majoron lifetime of $\tau\sim
10~\mathrm{sec}$ for the $Z_6\times Z_5$ model and $\tau\sim
10^5~\mathrm{sec}$ for the $Z_6\times Z_7$ model. Such a Majoron
can modify the big-bang nucleosynthesis processes. However the
modification is not significant since the Majoron will decouple
before the electro-weak phase transition. Its contribution to the
expansion rate is equivalent to that of $0.047\times 4/7\sim
0.027$ light neutrino species \cite{nucleonsythen}. This extra
contribution is well within observational uncertainties.

\section{Conclusions}

We have shown in this paper that the SM lagrangian including small
neutrino masses and with three generations has an anomaly-free
$Z_6$ discrete symmetry. This $Z_6$ can act as gauged baryon
parity and stabilize the nucleon even when the new physics
threshold is below the TeV scale. The $Z_6$ has a natural
embedding in $I^3_R+L_i+L_j-2L_k$ gauge symmetry. We have also
shown that realistic neutrino masses can be obtained through an
additional anomaly-free $Z_5$ or $Z_7$ symmetry which may have its
origin in the anomalous $U(1)_A$ symmetry.

Our framework can be readily extended to incorporate
supersymmetry. The $Z_6$ charge assignment is still as in Table
\ref{z6charge}. The down-type Higgs $\bar{H}$ has a $Z_6$ charge
of $5$ and the Grassmann variable $\theta$ has charge $3$ or $0$
\cite{wk}. The Higgsino contribution to the ${[SU(2)_L]}^2\times
Z_6$ anomaly is $3$ and the gaugino contribution to
${[SU(2)_L]}^2\times Z_6$ and ${[SU(3)_C]}^2\times Z_6$  anomalies
are $6$ and $9$ when $\theta$ charge is not zero. Since these
contributions are all multiples of $6/2$, the $Z_6$ symmetry
remains anomaly-free in supersymmetric case. This $Z_6$ forbids
all $\Delta B=1$ and $\Delta B=2$ nucleon decay as in the
non-supersymmetric case.

\section{Acknowledgement}

We are grateful to Yukihiro Mimura for useful discussion. This
work is supported in part by DOE Grant \# DE-FG03-98ER-41076, a
grant from the Research Corporation and by DOE Grant \#
DE-FG02-01ER-45684.

\end{document}